# Physical interpretation of fractional diffusion-wave equation via lossy media obeying frequency power law

W. Chen and S. Holm

Simula Research Laboratory, P. O. Box. 134, 1325 Lysaker, Norway

(15 March 2003)

The fractional diffusion-wave equation (FDWE)[1,2] is a recent generalization of diffusion and wave equations via time and space fractional derivatives. The equation underlies Levy random walk and fractional Brownian motion[2,3] and is foremost important in mathematical physics for such multidisciplinary applications as in finance, computational biology, acoustics, just to mention a few. Although the FDWE has been found to reflect anomalous energy dissipations[4,5], the physical significance of the equation has not been clearly explained in this regard. Here the attempt is made to interpret the FDWE via a new time-space fractional derivative wave equation which models frequency-dependent dissipations observed in such complex phenomena as acoustic wave propagating through human tissues, sediments, and rock layers. Meanwhile, we find a new bound (inequality (6) further below) on the orders of time and space derivatives of the FDWE, which indicates the so-called sub-diffusion process contradicts the real world frequency power law dissipation. This study also shows that the standard approach, albeit mathematically plausible, is physically inappropriate to derive the normal diffusion equation from the damped wave equation, also known as Telegrapher's equation.

The fractional diffusion wave equation[1,2] reads

$$\frac{\partial^{\beta} u}{\partial t^{\beta}} = -\kappa(-\Delta)^{\lambda/2} u, \quad 0 \prec \lambda, \beta \leq 2, \tag{1}$$

where $\Delta$ is the Laplacian operator, and $\kappa$ denotes a physical constant. $\lambda$ and $\beta$ can be arbitrary real number. For $\lambda=\beta=2$, equation (1) is the normal wave equation

$\partial^2 u/\partial t^2 = \kappa \Delta u$; for $\lambda=2$, $\beta=1$, it is the normal diffusion equation $\partial u/\partial t = \kappa \Delta u$. For non-integer $\lambda$ and $\beta$, equation (1) presents the fractional time derivative and the fractional Laplacian. For detailed discussions on the fractional derivative see ref. 6.

Table. I. Attenuation coefficients of frequency-dependent power law dissipation

|  | Water | Fat | Duct cancer | Body tissue |
| --- | --- | --- | --- | --- |
| $\alpha_0$ (dB/cm/MHz$^y$) | 0.0022 | 0.158 | 0.57 | 0.87 |
| $y$ | 2 | 1.7 | 1.3 | 1.5 |

- Boundary layer loss of rigid tube: $y=0.5$
- Underwater sediments and rock layers: $y \approx 1$

Essentially, equation (1) models non-conservative systems[4] and accounts for possible non-local and memory effects on energy dissipations[5], namely, the frequency-dependent dissipation $E = E_0 e^{-\alpha(\omega)z}$ observed, for example, in medical ultrasonic and seismic wave propagations[7-9]. Here $E$ represents the amplitude of an acoustic field variable such as velocity or pressure, and $\omega$ is angular frequency. The attenuation coefficient $\alpha(\omega)$ is characterized for a wide range of frequencies of practical interest by a power law function

$$\alpha(\omega) = \alpha_0 |\omega|^y, \qquad y \in [0,2], \qquad (2)$$

where $\alpha_0$ and $y$ are media-specific attenuation parameters obtained through a fitting of measured data. For most solid and highly viscous materials, $y$ is close to 2; while for some media of practical interest such as biomaterials, $y$ is from 1 to 1.7 (Table 1). For $y \neq 0,2$, the attenuation process can not be well described by common partial differential equation of integer order[7-9], and thus, is often called anomalous attenuation or diffusion. In recent decades, the fractional calculus has been found to be a powerful mathematical apparatus in modeling anomalous diffusion process[8-11]. However, unfortunately, the explicit relationship between the fractional diffusion wave equation (1) and the power law dissipation (2) has not clearly been unveiled. This is a major issue to be addressed in this study.

It is noted that in anomalous dissipation modeling, most effort has been concentrated on using the fractional time derivative[8,11]. Instead, very recently the present authors developed space fractional Laplacian lossy wave equation models[12]. By using both the fractional space/time derivatives, we have a new wave equation model for frequency dependent lossy media

$$\Delta p = \frac{1}{c_0^2}\frac{\partial^2 p}{\partial t^2} + \gamma \frac{\partial^\eta}{\partial t^\eta}(-\Delta)^{s/2} p,$$

$$0 \leq s \leq 2,\ 0 \prec \eta \leq 3,\ \eta \neq 2,\ 0 \leq y = s + \eta - 1 \leq 2, \tag{3}$$

where $\gamma$ is viscous constant, $s$ and $\eta$ can be arbitrary real number. By using time and space Fourier transforms[7], it is easy to verify that model equation (3) satisfies the power law (2). If $\Delta p$ is relatively small[7], the hyperbolic wave equation (3) can be approximated to the parabolic equation by removing $\Delta p$, namely,

$$\frac{1}{c_0^2}\frac{\partial^2 p}{\partial t^2} + \gamma \frac{\partial^\eta}{\partial t^\eta}(-\Delta)^{s/2} p = 0. \tag{4}$$

Integrating (4) with respect to time $t$ and multiplying by $c_0^2$ gives

$$\frac{\partial^{2-\eta} p}{\partial t^{2-\eta}} = -c^2 \gamma (-\Delta)^{s/2} p. \tag{5}$$

Compared with equation (1), one can see that equation (5) is the fractional diffusion-wave equation exhibiting frequency dependent dissipation obeying the power law (2), where the power coefficient $y=s+\eta-1$. Thus, we can interpret the fractional diffusion-wave equation (1) physically through the empirical power law frequency dissipation (2).

Comparing equations (1) with (5), it is straightforward that $y=s+\eta-1\in[0,2]$ in equation (5) leads to a bound on the derivative orders of fractional diffusion-wave equation (1)

$$-1\le \lambda - \beta \le 1. \tag{6}$$

To the best of the author's knowledge, inequality bound (6) is new. For $\lambda=2$, (6) requires $\beta\ge 1$. Therefore, the so-called sub-diffusion ($\lambda=2$, $0<\beta<1$)[1] does not agree with the power law (2), except of a negative exponent $y$ which indicates the inverse dependency of dissipation on frequency and is rarely, if not, found in the real world (in fact, vast majority falls in $1\le y\le 2$). On the other hand, the so-called super-diffusion process[1] ($\lambda=2$, $\beta>1$ or $0<\lambda<2$, $\beta=1$) satisfies inequality bound (6). The Baglegy-Torvik viscous equation[10] ($\lambda=0$, $\beta=0.5$) also meets (6).

Note that if $\eta>2$, the attenuation coefficient bound in equation (3) requires $s=0$, and the fractional diffusion equation (5) then degenerates into the reaction equation $\partial^{\eta-2}p/\partial t^{\eta-2} = -p/\gamma c^2$. On the other hand, for $\eta=1$, $s=0$, equation (3) turns out to be the damped wave equation, also known as Telegrapher's equation

$$\Delta p = \frac{1}{c_0^2}\frac{\partial^2 p}{\partial t^2} + \gamma \frac{\partial p}{\partial t}. \tag{7}$$

Assuming the velocity $c$ is very high, the second right-hand term can therefore be ignored and equation (7) then simplifies to the normal diffusion equation $\gamma\, \partial u/\partial t = \Delta u$. This is a standard approach to derive the parabolic diffusion equation from hyperbolic wave equation. It is well known that the normal diffusion equation exhibits the frequency-squared dependent dissipation[8,11,12]. However, in stark contrast, the Telegrapher's equation (7) itself describes frequency-independent dissipation[7]. Therefore, in the regime of the frequency dependent power, it is mathematically plausible but not physically sound to derive the normal diffusion equation from the Telegrapher's equation.

Considering the thermoviscous wave equation ($\eta=1$, $s=2$ in equation (3))[7]

$$\Delta p = \frac{1}{c_0^2}\frac{\partial^2 p}{\partial t^2} - \gamma\frac{\partial}{\partial t}\Delta p, \tag{8}$$

and assuming that $\Delta p$ is relatively small and can be neglected, equation (8) can then be reduced to the normal diffusion equation. It is noted that the thermoviscous wave equation conforms the normal diffusion equation in underlying the same square dependency of dissipation coefficient on frequency. Thus, this derivation is both mathematically and physically reasonable.

In the context of kinetic physics, equation (1) reflects the Levy stable process and fractional Brownian motion[2], where probability density function $u \geq 0$ requires $0 \prec \lambda \leq 2$. However, it is not clear if $\beta>0$ is physically necessary. For $\lambda, \beta<0$ and $\eta, s<0$, equations (1) and (3) are fractional integral equation.

As verified by Blackstock (1985), the nonlinear viscous wave equations also imply the frequency dependent dissipation. For instance, we can extend the fractional diffusion-wave equation (1) to the generalized fractional Bergers equation,

$$\frac{\partial^\beta u}{\partial t^\beta} + u \cdot \nabla u = -\kappa(-\Delta)^{\lambda/2} u,$$

$$0 \prec \lambda \leq 2, \quad 0 \prec \beta \prec 2, \quad -1 \leq \lambda - \beta \leq 1, \tag{9}$$

which has the frequency dependent power $y=\lambda-\beta+1$ in terms of power law (2).